\documentclass[aps,prl,twocolumn,superscriptaddress,floatfix]{revtex4-1}

\usepackage[utf8]{inputenc}
\usepackage{graphicx,xcolor}
\usepackage{amsmath}
\usepackage{etoolbox}

\begin{document}


\title{Two-fluid dynamics in driven YBa$_2$Cu$_3$O$_{6.48}$}


\author{A. Ribak}
\email[]{amit.ribak@mpsd.mpg.de}
\affiliation{Max Planck Institute for the Structure and Dynamics of Matter, 22761 Hamburg, Germany}

\author{M. Buzzi}
\affiliation{Max Planck Institute for the Structure and Dynamics of Matter, 22761 Hamburg, Germany}

\author{D. Nicoletti}
\affiliation{Max Planck Institute for the Structure and Dynamics of Matter, 22761 Hamburg, Germany}

\author{R. Singla}
\affiliation{Max Planck Institute for the Structure and Dynamics of Matter, 22761 Hamburg, Germany}

\author{Y. Liu}
\affiliation{Max Planck Institute for Solid State Research, 70569 Stuttgart, Germany}

\author{S. Nakata}
\affiliation{Max Planck Institute for Solid State Research, 70569 Stuttgart, Germany}

\author{B. Keimer}
\affiliation{Max Planck Institute for Solid State Research, 70569 Stuttgart, Germany}

\author{A. Cavalleri}
\email[]{andrea.cavalleri@mpsd.mpg.de}
\affiliation{Max Planck Institute for the Structure and Dynamics of Matter, 22761 Hamburg, Germany}
\affiliation{Department of Physics, Clarendon Laboratory, University of Oxford, Oxford OX1 3PU, United Kingdom}

\date{\today}

\begin{abstract}
Coherent optical excitation of certain phonon modes in YBa$_2$Cu$_3$O$_{6+x}$ has been shown to induce superconducting-like interlayer coherence at temperatures higher than $T_c$. Recent work has associated these phenomena to a parametric excitation and amplification of Josephson plasma polaritons, which are overdamped above $T_c$ but are made coherent by the phonon drive. However, the dissipative response of uncondensed quasiparticles, which do not couple in the same way to the phonon drive, has not been addressed. Here, we investigate both the enhancement of the superfluid density, $\omega\sigma_2(\omega)$, and the dissipative response of quasiparticles, $\sigma_1(\omega)$, by systematically tuning the duration and energy of the mid-infrared pulse while keeping the peak field fixed. We find that the photo-induced superfluid density saturates to the zero-temperature equilibrium value for pulses made longer than the phonon dephasing time, whilst the dissipative component continues to grow with increasing pulse duration. We show that superfluid and dissipation remain uncoupled as long as the drive is on, and identify an optimal regime of pump pulse durations for which the superconducting response is maximum and dissipation is minimized.
\end{abstract}


\maketitle

When cooled below the equilibrium superconducting transition, cuprate superconductors acquire a characteristic terahertz-frequency optical response, reflecting coherent interlayer transport and condensation of normal state quasiparticles \cite{ref1,ref2,ref3,ref4}. These features are displayed in Fig. 1(a,b), where we report the complex optical conductivity of underdoped YBa$_2$Cu$_3$O$_{6+x}$ along the $c$-axis for $T > T_c$ (red) and $T < T_c$ (blue). In the normal state, the real (dissipative) part of the conductivity, $\sigma_1(\omega)$, at low frequencies ($\omega \leq 100 \ \text{cm}^{-1}$) is finite, exhibiting a flat and featureless spectrum (red curve in Fig. 1(a), left panel). For $\omega > 100$ cm$^{-1}$ (Fig. 1(a), right panel) a collection of narrow absorption peaks is observed, assigned to infrared-active phonons, along with a broader contribution around 400 cm$^{-1}$, which has been attributed to the so-called “transverse Josephson plasmon” \cite{ref5,ref6,ref7,ref8}. Notably, unlike in conventional superconductors, the spectral weight loss across the superconducting transition (blue shading) extends over a frequency range which is far broader than the superconducting gap, a behavior which is commonly found in strongly-correlated superconductors \cite{ref1,ref2,ref3,ref4,ref9}. The presence of a zero-frequency pole in $\sigma_1(\omega)$ manifests itself in the imaginary (inductive) part of the optical conductivity, $\sigma_2(\omega)$, as a $1/\omega$ divergence at low frequency (Fig. 1(b)), which allows the determination of the superfluid density from finite frequency measurements, as $\lim_{\omega \to 0} \omega\sigma_2(\omega)=\frac{1}{\omega}\frac{n_se^2}{m}$ (here $n_s$ is the superfluid density, $e$ the electron charge and $m$ the electron mass).

In a recent series of experiments, femtosecond mid-infrared pulses have been used to resonantly excite apical oxygen vibrations in underdoped YBa$_2$Cu$_3$O$_{6+x}$ \cite{ref10,ref11,ref12,ref13,ref14,ref15,ref16}. These studies revealed a light-induced transient optical response featuring a finite light induced superfluid density, as seen most directly in the transient $\sigma_2(\omega)$ which acquires the same $1/\omega$ behavior observed below $T_c$. Strikingly, these effects were observed all the way up to the pseudo-gap temperature, $T^*$ \cite{ref10,ref11,ref12,ref13}. A representative result is reported in Fig. 1(c,d), where we show the complex optical conductivity of YBa$_2$Cu$_3$O$_{6.48}$ measured at $T = \text{100 K} \approx 2 T_c$ before (red lines) and after (blue circles) mid-infrared photoexcitation (here, the region covered by the pump spectrum is shaded in blue). Notably, this analogy does not apply to the real part of optical conductivity. In the transient state the low-frequency $\sigma_1(\omega)$ shows an increase (Fig. 1(c), left panel), as opposed to a decrease upon cooling (Fig. 1(a), left panel). This dissipative response can be reproduced by fitting the data with a two-fluid model in which a superconducting term, given by a Josephson Plasma Resonance, coexists with an overdamped Drude absorption \cite{ref12,ref17,ref18,ref19} that accounts for the tunnelling of incoherent normal carriers (dashed lines in Fig. 1(c,d), see Supplemental Material for details on the fitting model).

Recently, further experimental and theoretical work \cite{ref20,ref21} has provided evidence that periodic driving of the apical oxygen modes acts as a parametric pump for the amplification of incoherent Josephson plasmons in the pseudogap phase, giving rise to the superconducting-like optical properties observed with THz spectroscopy. In this context, it is not clear how sustained drives affects the dissipative response, and how the ratio of superfluid density to dissipative heating can be optimized. 

\begin{figure*}[htbp]
	\includegraphics[width=2\columnwidth]{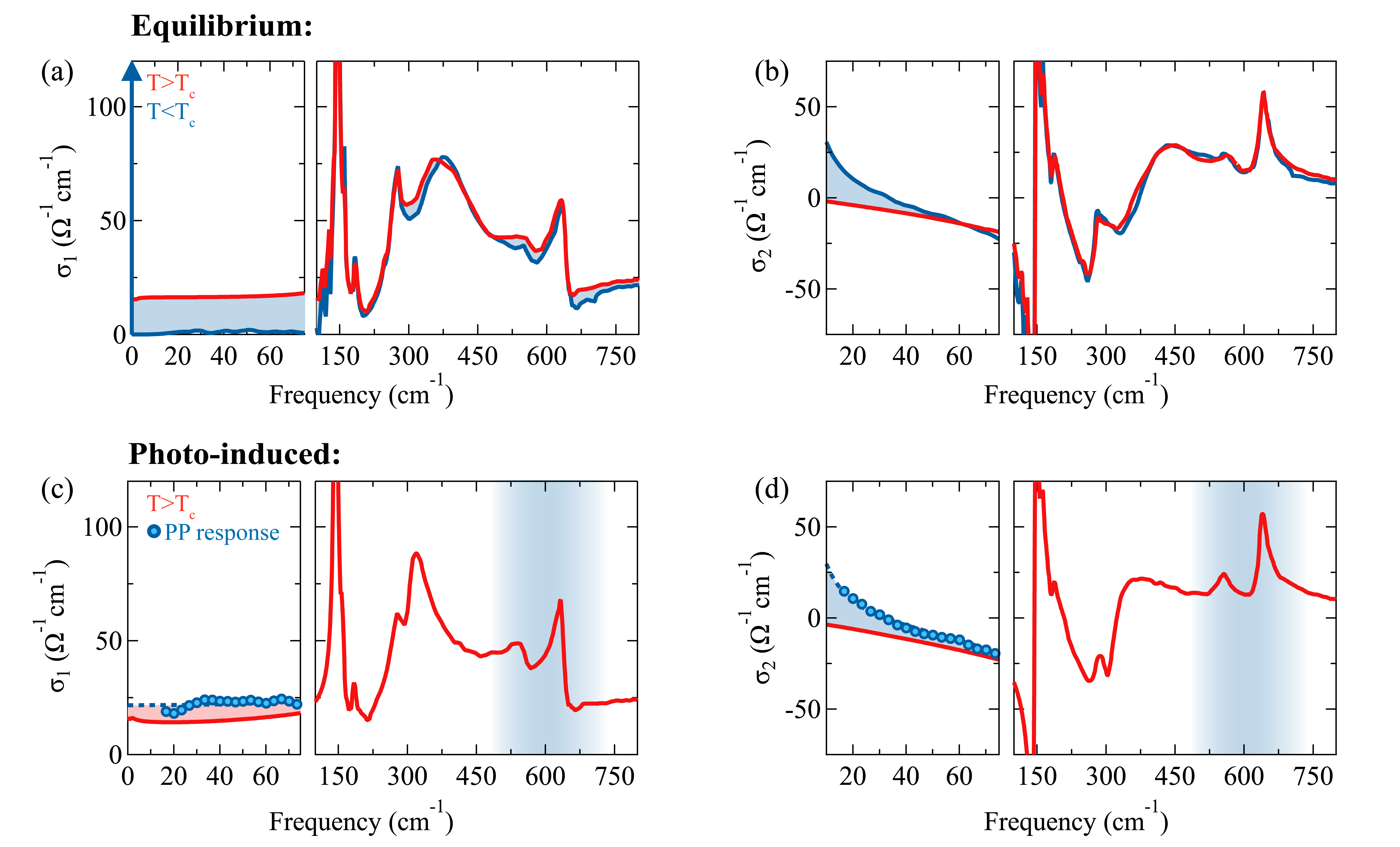}
    \caption{(a,b) Complex $c$-axis optical conductivity, $\sigma_1(\omega)+i\sigma_2(\omega)$, of underdoped YBa$_2$Cu$_3$O$_{6+x}$ across the equilibrium superconducting transition. Data at $T = 60 \text{ K} > T_c$ (red) and $T = 10 \text{ K} < T_c$ ($T_c=52$ K) are reported \cite{ref2,ref11}. Filled blue regions represent the changes in the spectra across $T_c$. The vertical arrow in (a) is a pole at zero frequency indicative of dissipationless transport.  (c,d) Same quantities as in (a,b) measured in YBa$_2$Cu$_3$O$_{6.48}$ at $T=100$ K  before (red) and after (blue circles) mid-infrared photo-excitation (see main text). Red and blue fillings in (c) and (d) represent photo-induced changes in the spectra, while the shaded blue area around 600 cm$^{-1}$ indicates the frequency range covered by the mid-infrared pump. Dashed blue lines are fits to the transient optical conductivity performed with a model including a Josephson plasmon and a Drude term (see also Supplemental Material).\label{fig:Fig1}}
\end{figure*}

Here, we explore this issue by photo-exciting YBa$_2$Cu$_3$O$_{6.48}$ with mid-infrared pump pulses of variable duration, tuned to resonantly drive the same $\text{570 cm}^{-1}$ and $\text{670 cm}^{-1}$ apical oxygen phonons that induce the superconducting-like behavior discussed above \cite{ref11,ref12}. To this end, single crystals of YBa$_2$Cu$_3$O$_{6.48}$ with typical dimensions of $1.5 \times 1.5 \times 0.4$ $\text{mm}^3$ were mounted to expose a surface determined by the out-of-plane ($\parallel c$) and one of the in-plane ($\perp c$) crystallographic directions. The crystals were photo-excited using mid-infrared pump pulses generated using an optical parametric amplifier pumped by a commercial Ti:Sa laser system. These pump pulses were polarized along the $c$ axis and made to propagate through dispersive, optically polished NaCl plates before impinging on the sample. This yielded excitation pulses with the same spectral content but duration varying from $\text{150 fs}$ up to $\text{4.3 ps}$. Broadband THz probe pulses ($\approx \text{17} - \text{75 cm}^{-1}$) were generated using a photoconductive antenna from the direct output of the Ti:Sa amplifier. These THz probe pulses, with polarization parallel to the crystal $c$ axis, were then focused on the sample and detected by electro-optical sampling after reflection in a 500 $\mu$m thick ZnTe (110) crystal, yielding the photo-induced changes in the low frequency complex reflection coefficient $\tilde{r}(\omega)$ as a function of pump-probe time delay. The penetration depth of the mid-infrared pump ($\approx$ 1 $\mu$m) was shorter than that of THz probe (5-10 $\mu$m). This was taken into account by modelling the sample as a multi-layered photo-excited stack on top of an unperturbed bulk in order to obtain the optical response functions corresponding to an effective semi-infinite and homogeneously excited medium (see Supplemental Material and Ref. \cite{ref22}).

The measurements presented in the following were carried out at a constant peak field of $\text{2.2 MV/cm}$ for all pulse durations. In the scenario of parametrically amplified superconductivity \cite{ref20,ref21,ref22,ref23,ref24}, the use of sustained drives at constant peak field allows us to maintain the amplitude of the parametric drive constant for increasing pulse duration, thus enabling the study of the interplay between dissipation and superconductivity in the photo-induced state.

\begin{figure}[htbp]
	\includegraphics[width=1\columnwidth]{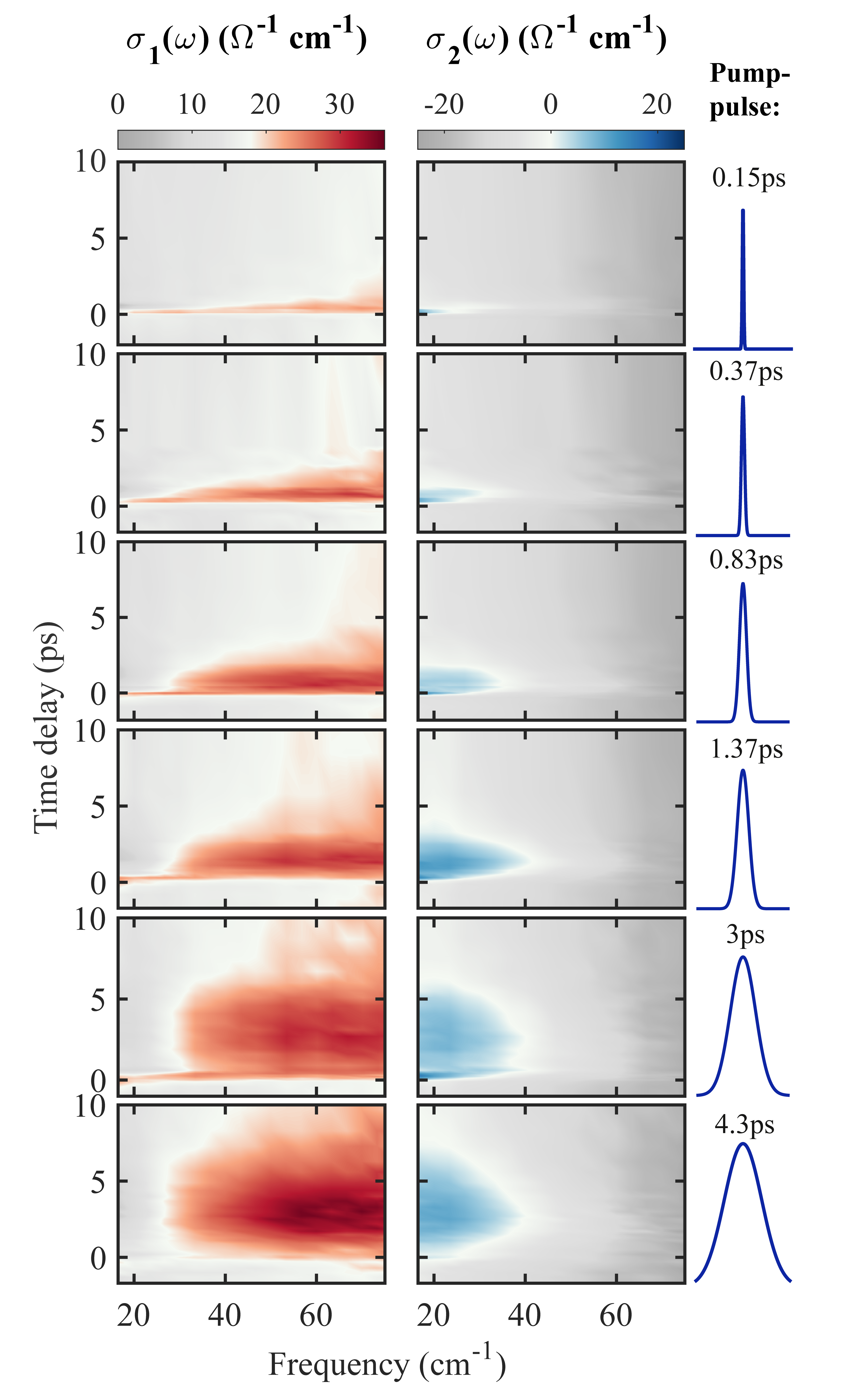}
    \caption{Frequency- and time-delay-dependent complex optical conductivity of YBa$_2$Cu$_3$O$_{6.48}$ after photo-excitation with different mid-infrared pump pulse duration at $T=100$ K. The peak field was kept constant at 2.2 MV/cm for all measurements.\label{fig:Fig2}}
\end{figure}

Figure 2 provides a visual summary of the results of the pulse duration dependence, for which we report extended data sets in the Supplemental Material. The color plots show the frequency- and time-dependent transient complex optical conductivity $\sigma_1(\omega)+i\sigma_2(\omega)$ after excitation with different mid-infrared pulse durations varying from $\text{0.15 ps}$ to $\text{4.3 ps}$ at a temperature $T=100$ K $\gg T_c$ ($\cong$ 52 K). The $\sigma_1(\omega)$ spectra show an almost frequency independent increase (red colors) at all positive time delays which is indicative of enhanced dissipation. The imaginary conductivity, $\sigma_2(\omega)$, displays instead a $1/\omega$-like divergence at low frequency (blue shaded areas), which persist for a time window of the order of the pump pulse duration. Besides the new observation of an increased lifetime of the transient superconducting response, our data are consistent with the previous findings reported in Refs. \cite{ref10,ref11,ref12}. However, while the qualitative features in the optical spectra do not depend on the drive pulse duration, we find quantitative differences, analyzed in the following.

To reconstruct the dynamical evolution of the two components of the transient response, namely the dissipative one in $\sigma_1(\omega)$ and the coherent one in $\sigma_2(\omega)$, we extracted two representative quantities from the individual THz spectra measured at each time delay. 

The transient $\sigma_1(\omega)$ spectral weight, indicative of dissipation, is determined as $\int_{20\text{ cm}^{-1}}^{75\text{ cm}^{-1}}\sigma_1(\omega) \mathrm{d}\omega$, 
while the coherent response is expressed as $\lim_{\omega \to 0} \omega\sigma_2(\omega)$, which in a superconductor at equilibrium is proportional to the superfluid density. Figures 3(a,b) and 3(c,d) display these two figures of merit for representative pulse durations of 0.15 ps and 3 ps, respectively. For both data sets, after photo-excitation the $\sigma_1(\omega)$ spectral weight is enhanced and then relaxes on a time scale that clearly exceeds the pump pulse duration. A different evolution is observed instead in  $\lim_{\omega \to 0} \omega\sigma_2(\omega)$, 
which appears to be finite only while the system is driven (blue shading in Fig. 3(c,d)). In this case, excitation with longer pulses allows the “zero-temperature” equilibrium superfluid density to be reached. This is reported for comparison as a dashed horizontal line in Fig. 3(d). A more accurate visualization of how the lifetime of the photoinduced superconducting state scales with the pump pulse duration is displayed in Fig. 3(e), emphasizing a one-to-one correspondence between drive duration and lifetime.

\begin{figure*}[htbp]
	\includegraphics[width=2\columnwidth]{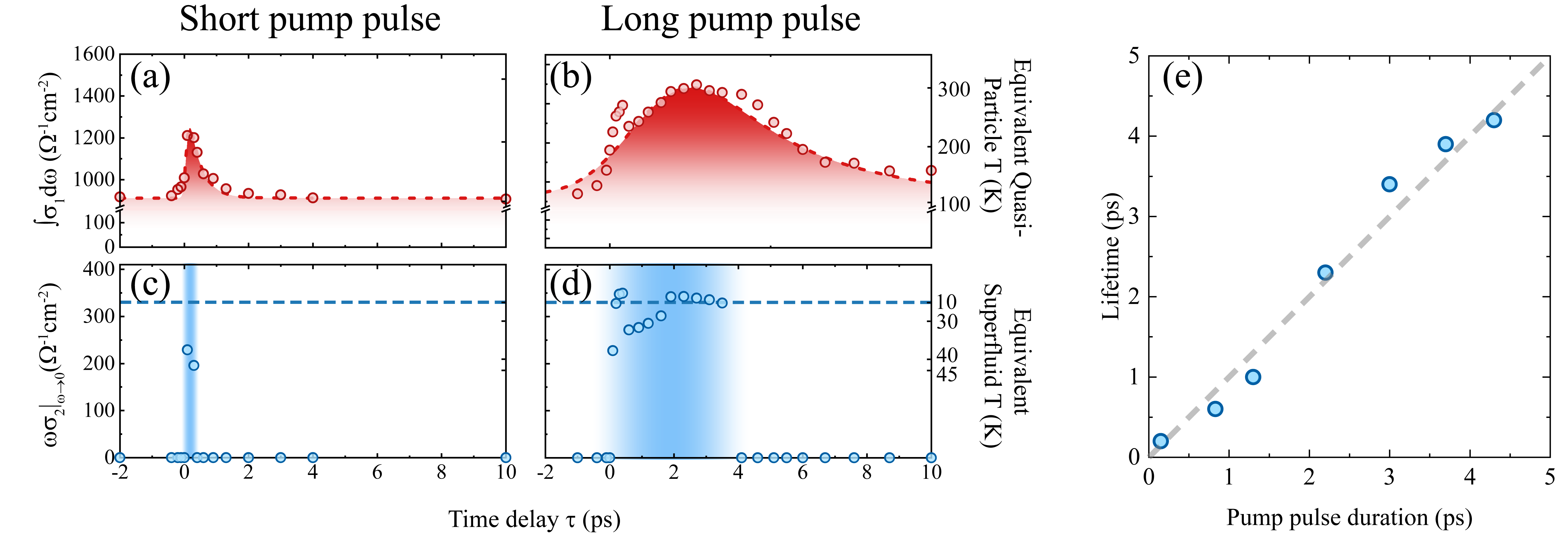}
    \caption{(a,b) Dynamical evolution of the transient spectral weight, $\int_{20\text{ cm}^{-1}}^{75\text{ cm}^{-1}}\sigma_1(\omega) \mathrm{d}\omega$, indicative of dissipation, as a function of pump-probe time delay $\tau$, for unstretched (0.15 ps) and stretched (3 ps) excitation, respectively. The dashed line is a fit to the data with a finite rise time and an exponential decay. (c,d) Corresponding evolution of the coherent superconducting-like response, $\lim_{\omega \to 0} \omega\sigma_2(\omega)$. The horizontal lines indicate the 10 K equilibrium value of the superfluid density, while the shaded area represent the time delay window for which a finite coherent response is detected. Equivalent quasi-temperatures (see main text) are reported on the right axes. (e) Lifetime of the photo-induced superconducting state, estimated from the width of the shaded regions in (c,d), as a function of pump pulse duration.\label{fig:Fig3}}
\end{figure*}

The transient $\sigma_1(\omega)$ spectral weight and  $\lim_{\omega \to 0} \omega\sigma_2(\omega)$ can be compared to their respective values in equilibrium at different temperatures to extract a “quasi-temperature” of these two fluids, which we also provide in Figs. 3(a,b) and 3(c,d). Unlike in the equilibrium superconductor, in this driven state some normal quasiparticles remain and are heated up by the pump up to an apparent temperature of $\approx 300$ K, while the parametric drive appears to synchronize pre-existing Cooper pairs yielding an effective superfluid quasi-temperature of $\approx 10$ K.

In Figure 4 we display the same quantities as in Fig. 3(a-d) as a function of pump pulse duration, measured at the peak of the response, in order to study how the dissipative and superfluid components are affected by prolonged driving. Our data reveal that, while the normal dissipative part increases monotonously with the pump duration (Fig. 4(a)) with no sign of saturation, the coherent superconducting part (Fig. 4(b)) saturates almost exactly at the equilibrium value of the superfluid density measured below $T_c$, at a pump pulse duration of $\approx \text{1 ps}$. This value matches closely the lifetime of the resonantly excited apical oxygen vibration, as estimated from both the linewidth of the absorption peak in the conductivity spectrum \cite{ref3} and the decay time of coherently driven oscillations in the mid-infrared pump / second harmonic probe experiment of Ref. \cite{ref20}.

\begin{figure}[htbp]
	\includegraphics[width=1\columnwidth]{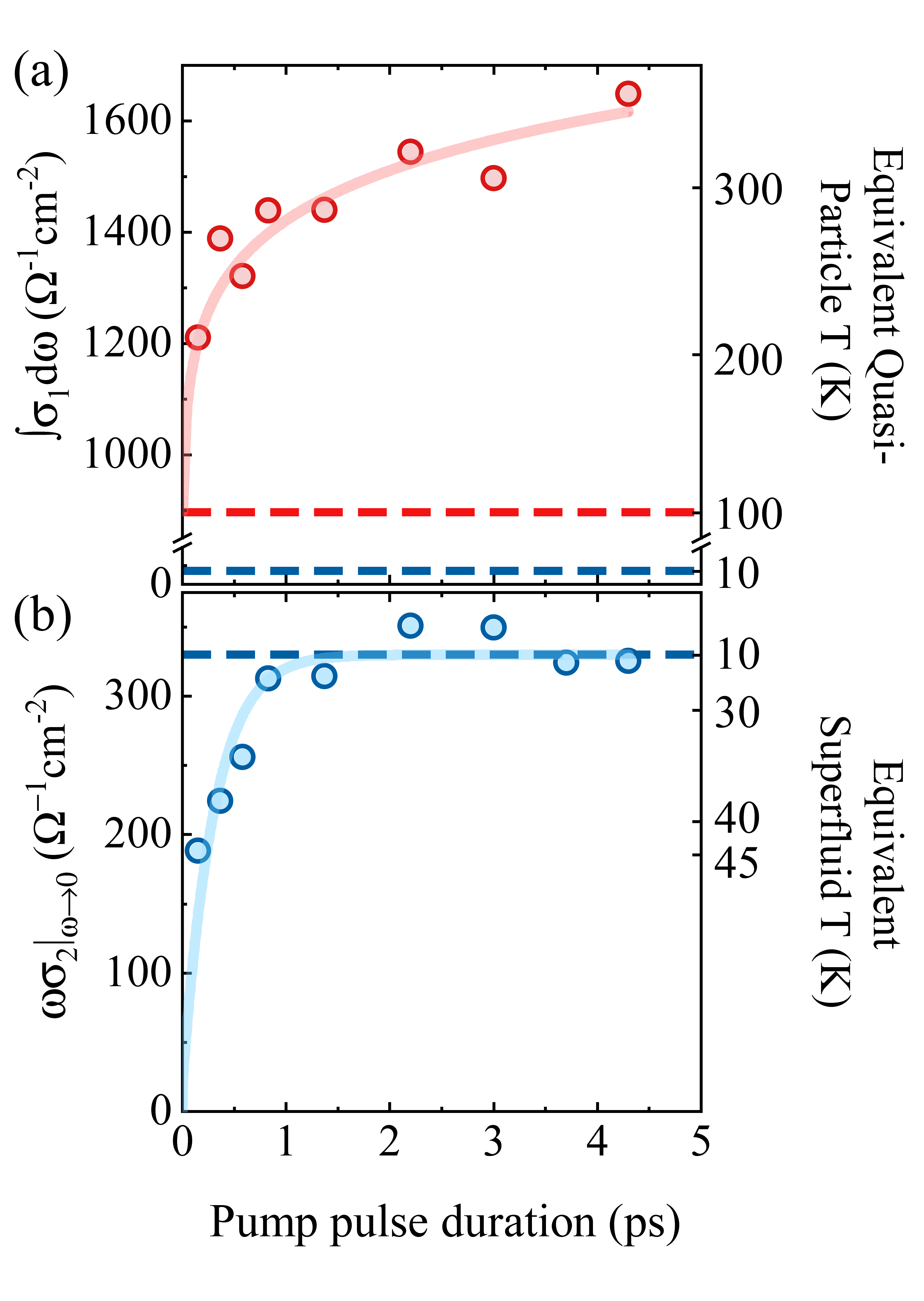}
    \caption{(a) Evolution of the transient spectral weight, $\int_{20\text{ cm}^{-1}}^{75\text{ cm}^{-1}}\sigma_1(\omega) \mathrm{d}\omega$, and (b) of the coherent superconducting-like response, $\lim_{\omega \to 0} \omega\sigma_2(\omega)$, measured at the time delay corresponding to the peak of the response, as a function pump pulse duration. The red and blue horizontal dashed lines in (a) indicate the equilibrium spectral weight values at $T=100$ K and $T=10$ K, respectively, while that in (b) refers to the equilibrium superfluid density at $T=10$ K. Equivalent temperatures are reported on the right axes (see discussion in main text).\label{fig:Fig4}}
\end{figure}

Our observations show how the dynamics of synchronized superfluid and heated quasiparticles is uncorrelated as long as the drive is on. We speculate that dissipation occurs along the nodal directions of the Brillouin zone while the coherent response involves only the anti-nodal areas of the pseudogap. This interpretation assumes preexisting electron pairs without phase coherence. The possible creation of new pairs appears unlikely, as the transient superfluid density matches the value at equilibrium.

Finally, our measurement allows us to identify an ideal excitation regime, one with drive pulses of the order of $\text{1 ps}$, for which the superconducting-like response reaches the “zero-temperature” value of the superfluid density and is maximized in comparison to the dissipative component, thus indicating a path for optimizing light-induced superconductivity in bilayer cuprates.

\begin{acknowledgments}
The research leading to these results received funding from the European Research Council under the European Union’s Seventh Framework Programme (FP7/2007-2013)/ERC Grant Agreement No. 319286 (QMAC). We acknowledge support from the Deutsche Forschungsgemeinschaft (DFG, German Research Foundation) via the excellence cluster ‘The Hamburg Centre for Ultrafast Imaging’ (EXC 1074 – project ID 194651731) and the priority program SFB925 (project ID 170620586).
\end{acknowledgments}

\end{document}



\title{\normalfont{Supplemental Material} \\
\bf{Two-fluid dynamics in driven YBa$_2$Cu$_3$O$_{6.48}$}}


\author{A. Ribak}
\email[]{amit.ribak@mpsd.mpg.de}
\affiliation{Max Planck Institute for the Structure and Dynamics of Matter, 22761 Hamburg, Germany}

\author{M. Buzzi}
\affiliation{Max Planck Institute for the Structure and Dynamics of Matter, 22761 Hamburg, Germany}

\author{D. Nicoletti}
\affiliation{Max Planck Institute for the Structure and Dynamics of Matter, 22761 Hamburg, Germany}

\author{R. Singla}
\affiliation{Max Planck Institute for the Structure and Dynamics of Matter, 22761 Hamburg, Germany}

\author{Y. Liu}
\affiliation{Max Planck Institute for Solid State Research, 70569 Stuttgart, Germany}

\author{S. Nakata}
\affiliation{Max Planck Institute for Solid State Research, 70569 Stuttgart, Germany}

\author{B. Keimer}
\affiliation{Max Planck Institute for Solid State Research, 70569 Stuttgart, Germany}

\author{A. Cavalleri}
\email[]{andrea.cavalleri@mpsd.mpg.de}
\affiliation{Max Planck Institute for the Structure and Dynamics of Matter, 22761 Hamburg, Germany}
\affiliation{Department of Physics, Clarendon Laboratory, University of Oxford, Oxford OX1 3PU, United Kingdom}

\date{\today}


\maketitle

\section{S1. Sample growth and characterization}

YBa$_2$Cu$_3$O$_{6.48}$ crystals of typical dimensions 1.5 x 1.5 x 0.4 mm were grown in Y-stabilized zirconium crucibles. The hole doping of the Cu-O planes was adjusted by controlling the oxygen content of the Cu-O chain layer by annealing in flowing O$_2$ and subsequent rapid quenching. A sharp superconducting transition at $T_c = 55$ K ($\Delta T_c \approx 2$ K) was determined by dc magnetization measurements, as shown in Fig. S1.

\section{S2. Experimental setup and data-acquisition}
To retrieve the photo-induced changes in the optical properties of YBa$_2$Cu$_3$O$_{6.48}$ we performed a series of mid-infrared pump / THz probe experiments with different pump pulse durations. The YBa$_2$Cu$_3$O$_{6.48}$ crystals were manually polished in order to expose a surface which contained both an in-plane and an out-of-plane axis. They were glued on top of cutting-edge shaped holders to prevent stray reflections from interfering with the signal from the sample surface. These holders were installed on the cold-finger of a He-flow cryostat to allow for temperature control.

\begin{figure}[b]
	\includegraphics[width=0.6\columnwidth]{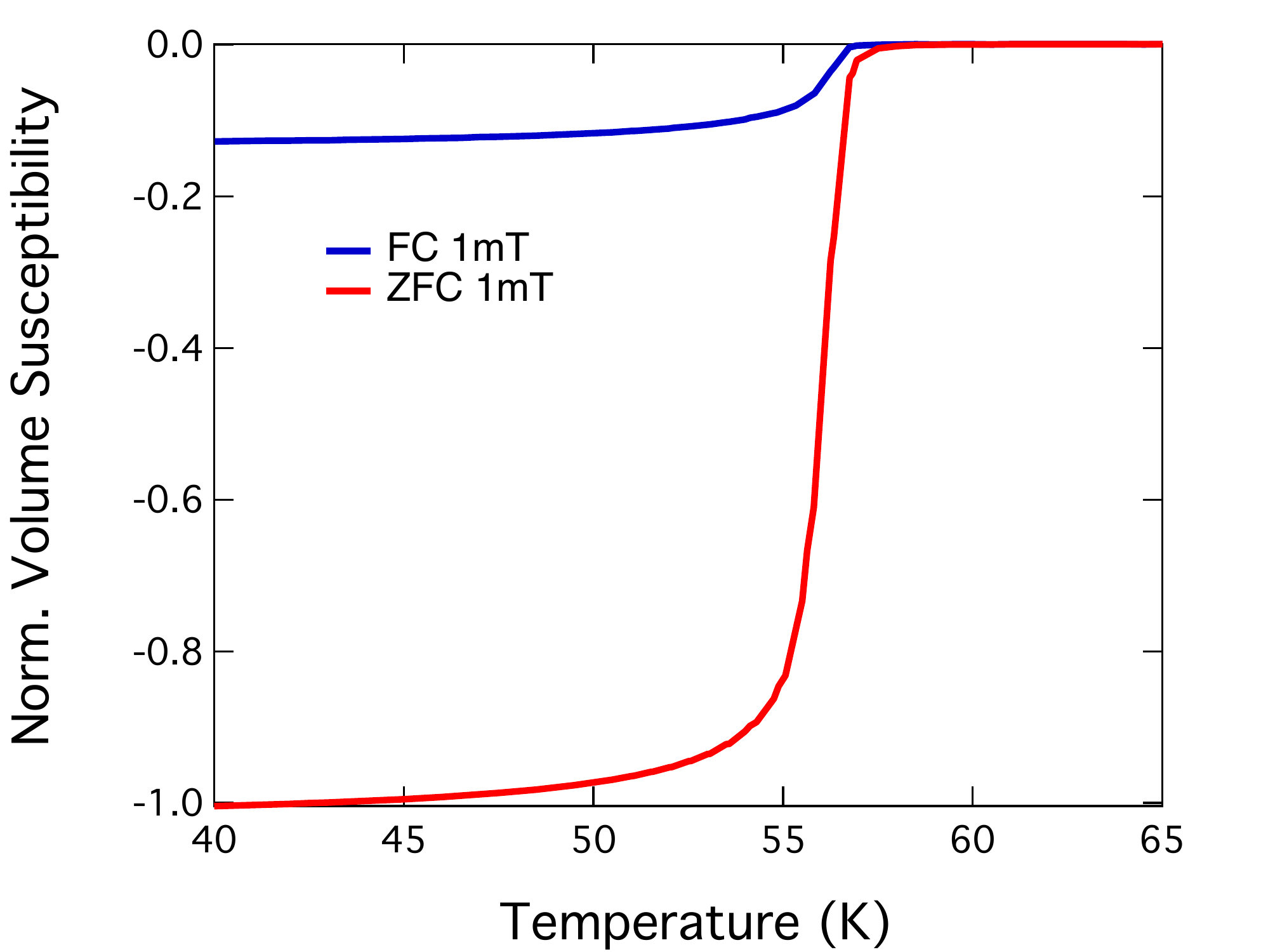}
    \caption{SQuID characterization of the DC magnetization (ZFC: zero field cooled, FC: field cooled) across the superconducting transition.\label{fig:FigS1}}
\end{figure}

\begin{figure}[b]
	\includegraphics[width=0.6\columnwidth]{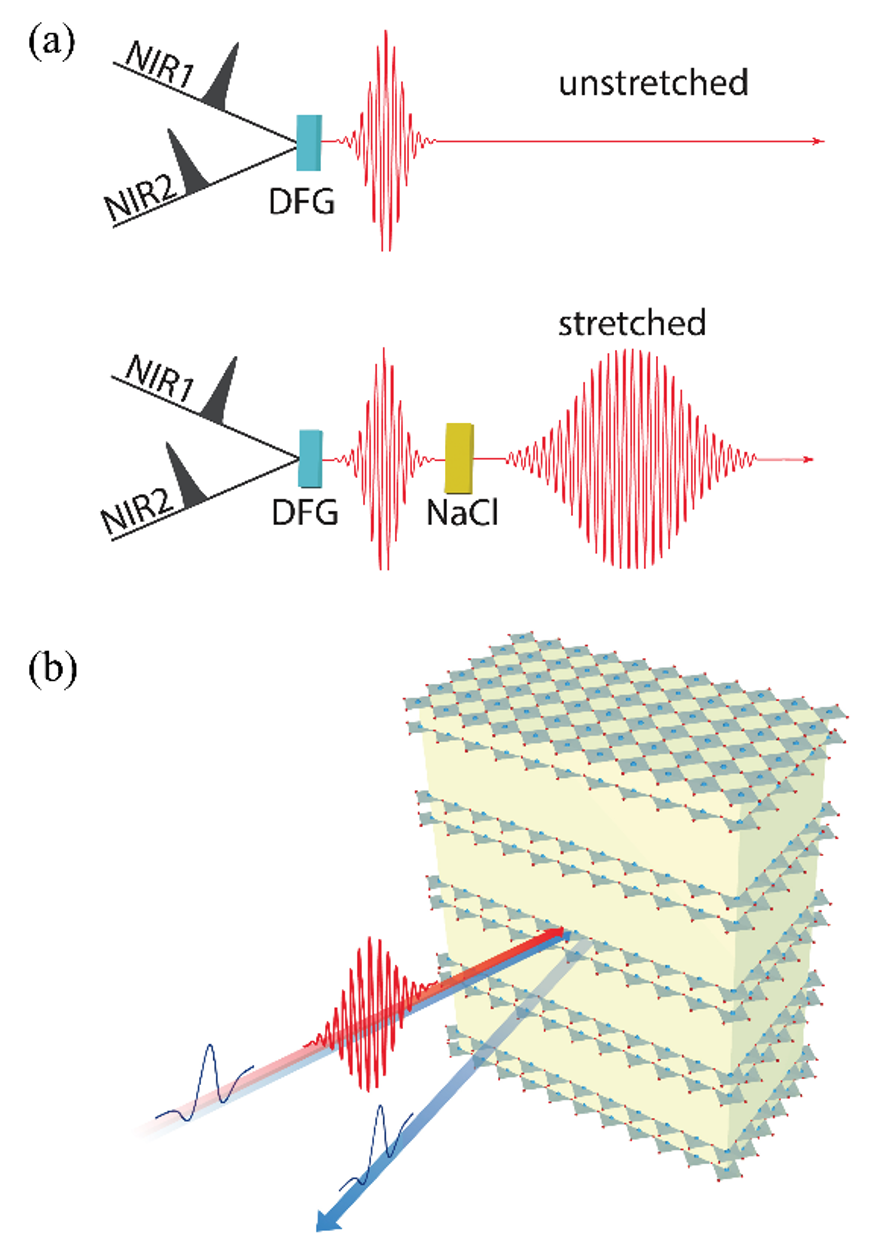}
    \caption{Pump-probe experimental setup. (a) The signal and idler outputs from an optical parametric amplifier are sent to a GaSe crystal for difference frequency generation (DFG). The duration of these pulses can be controlled by letting them propagate through optically polished NaCl plates. (b) The MIR pump pulse (red) and the single-cycle THz probe (blue) are polarized along the $c$ axis of YBa$_2$Cu$_3$O$_{6.48}$.\label{fig:FigS2}}
\end{figure}

YBa$_2$Cu$_3$O$_{6.48}$ was photo-excited with mid-infrared pump pulses of duration variable between 150 fs and 4.3 ps, tuned to $\sim 20$ THz center frequency, in resonance to the $B_{1u}$ phonon modes that were previously shown to enhance interlayer superconducting coupling. 

These pump pulses were generated by difference frequency mixing of the signal and idler outputs of an optical parametric amplifier (OPA) in a 0.4-mm thick GaSe crystal which yielded 20-THz-center-wavelength pulses with total energy of $\sim 50$ $\mu$J and typical duration of ~150 fs. The OPA was pumped with $\sim 35$ fs long pulses from a commercial Ti:Sapphire regenerative amplifier (800 nm wavelength). The duration of the pump pulses was controlled by introducing dispersion through propagation in NaCl rods and their intensity was adjusted using two freestanding wire-grid polarizers to keep the peak electric field constant at 2.2 MV/cm.

The pump pulses were then focused onto the sample surface, with their polarization aligned with the $c$ axis of the YBa$_2$Cu$_3$O$_{6.48}$ crystal. The typical pump spot size was $\sim 0.5$ mm allowing for a homogenous illumination of the probed area. A maximum fluence of $\sim 30$ mJ/cm$^2$ could be achieved, allowing to maintain a 2.2 MV/cm peak field at the longest pulse duration of 4.3 ps.

The transient reflectivity changes after photo-excitation were retrieved via time-domain THz spectroscopy using single cycle THz pulses generated by illuminating a commercial photoconductive antenna with 800-nm pulses. These probe pulses, polarized along the $c$ axis of the YBa$_2$Cu$_3$O$_{6.48}$ crystal, were focused at normal incidence onto the sample surface on a spot of $\lesssim 0.3$ mm diameter. After reflection from the sample surface, their electric field was detected via electro-optic sampling in a 500-$\mu$m thick optically contacted ZnTe crystal. This setup provided a spectral bandwidth extending from $\sim 17$ to 75 cm$^{-1}$.

In order to minimize the effects on the pump-probe time resolution due to the finite duration of the THz probe pulse, we performed the experiment as described in Ref. \cite{Ref1}. 
The transient reflected field at each time delay $\tau$ after excitation was obtained by keeping fixed the delay $\tau$ between the pump pulse and the electro-optic sampling gate pulse, while scanning the delay $t$ of the single-cycle THz probe pulse. The stationary probe electric field $E_{R}(t)$ and the differential electric field $\Delta E_{R}(t,\tau)$ reflected from the sample were recorded simultaneously by feeding the electro-optic sampling signal into a digitizer that sampled individual laser pulses, and mechanically chopping the pump and probe beams at different frequencies. $E_{R}(t)$ and $\Delta E_{R}(t,\tau)$ were then independently Fourier transformed to obtain the complex-valued, frequency-dependent $\tilde{E}_{R}(\omega)$ and $\Delta \tilde{E}_{R}(\omega,\tau)$.

\section{S3. Determination of the transient optical properties}

From the frequency dependent measured quantities $\tilde{E}_{R}(\omega)$ and $\Delta \tilde{E}_{R}(\omega,\tau)$ the complex reflection coefficient in the photoexcited state could be extracted by inverting the following equation:

\begin{equation}
\frac{\Delta \tilde{E}_{R}(\omega,\tau)}{\tilde{E}_{R}(\omega)} = \frac{\tilde{r}_{pumped}(\omega,\tau)-\tilde{r}_{0}(\omega)}{\tilde{r}_{0}(\omega)}
\end{equation}
\\
The penetration depth ($\sim0.7$ $\mu$m) of the excitation pulses (defined as $d(\omega)=\frac{c}{2\omega} \text{Im}[\tilde{n}_{0}(\omega)]$, where $\tilde{n}_{0}(\omega)$ is the equilibrium complex refractive index) was typically smaller than that of the probe pulses (5-10 $\mu$m). Due to this, the probed volume was not homogenously excited and this was taken into account using the following approach.

As the pump penetrates the material, its intensity is reduced and will induce progressively weaker changes in the refractive index of the sample. We modelled this condition by “slicing” the probed thickness of the material into thin layers, where we assumed that the pump-induced changes in the refractive index were proportional to the pump intensity in the layer, i.e., $\tilde{n}(\omega,z)=\tilde{n}_{0}(\omega)+\Delta\tilde{n}(\omega)e^{-\alpha z}$. Here, $\tilde{n}_{0}(\omega)$ is the equilibrium complex refractive index, $\alpha$ is the attenuation coefficient at the pump frequency, and $z$ is the spatial coordinate along the sample thickness.

For each probe frequency $\omega$, the complex reflection coefficient $\tilde{r}(\Delta\tilde{n})$ of such a multilayer stack was calculated with a characteristic matrix approach \cite{Ref2}, keeping $\Delta\tilde{n}$ as a free parameter. Equation 1 relates the measured quantity $\frac{\Delta \tilde{E}_{R}(\omega,\tau)}{\tilde{E}_{R}(\omega)}$ to the changes in the complex reflection coefficient. One can extract $\Delta\tilde{n}$ by minimizing numerically:

\begin{equation}
\left|\frac{\Delta\tilde{E}_{R}(\omega,\tau)}{\tilde{E}_{R}(\omega)} - \frac{\tilde{r}(\omega,\Delta\tilde{n})-\tilde{r}_{0}(\omega)}{\tilde{r}_{0}(\omega)}\right|
\end{equation}
\\
The value of $\Delta\tilde{n}(\omega)$ retrieved by this minimization represents the pump-induced change in the refractive index at the surface. By taking $\tilde{n}(\omega)=\tilde{n}_{0}(\omega)+\Delta\tilde{n}(\omega)$, one can reconstruct the optical response functions of the material as if it had been homogeneously transformed by the pump.

\section{S4. Data-analysis and fitting models}

At each pump-probe time delay $\tau$, the transient $c$-axis optical response functions were fitted with a two-fluid model including the response of a Josephson plasma and a Drude contribution, to account for dissipation arising from residual quasiparticles \cite{Ref3,Ref4}. In addition to this, the phonon modes in the mid-infrared (150 cm$^{-1}$ $\lesssim\omega\lesssim$ 700 cm$^{-1}$) and the high-frequency electronic absorption ($\omega\gtrsim$ 3000 cm$^{-1}$) were modeled with a series of Lorentz oscillators, for which the complex dielectric function is expressed as: 

\begin{equation}
\tilde{\epsilon}_{HF}(\omega)=\sum_{i} \frac{\Omega^{2}_{p,i}}{(\Omega^{2}_{0,i}-\omega^{2})-i\omega\Gamma_i}
\end{equation}
\\
Here, $\Omega_{0,i}$, $\Omega_{p,i}$ and $\Gamma_{i}$ are the peak frequency, plasma frequency, and damping coefficient of the $i$-th oscillator, respectively.

The dielectric function of the two-fluid model is expressed as:

\begin{equation}
\tilde{\epsilon}_{LF}(\omega)=\epsilon_{\infty} - \frac{\omega^{2}_{J}}{\omega^2} - \frac{\omega^{2}_{D}}{\omega^{2}+i\Gamma_D \omega}
\end{equation}
\\
Here, the free fit parameters are the Josephson plasma frequency, $\omega_{J}$, which accounts for the superconducting-like response, and the Drude parameters $\omega_{D}$ and $\Gamma_{D}$, which are plasma frequency and scattering rate of the “normal” component. The value of $\epsilon_{\infty}$ was set to 4.5, according to previous literature \cite{Ref5}.

In all fitting procedures a single set of fit parameters was always used to simultaneously reproduce the real and imaginary part of the optical conductivity, $\sigma_1(\omega)$ and $\sigma_2(\omega)$. The equilibrium spectra above $T_c$ were modeled by keeping $\omega_{J}=$ 0, since no superfluid is present, letting all the other parameters free to vary. The result of these fits was used as a starting point for the transient data. In this procedure, the high-frequency terms contained in $\tilde{\epsilon}_{HF}$ were kept fixed while $\omega_{J}$, $\omega_{D}$, and $\Gamma_{D}$ were left free to vary.

\section{S5. Extended data sets}

In this Section we report an extended data set from which the graphs in Fig. 2 and Fig. 3 of the main text were produced.

\begin{figure}[h]
	\includegraphics[width=0.7\columnwidth]{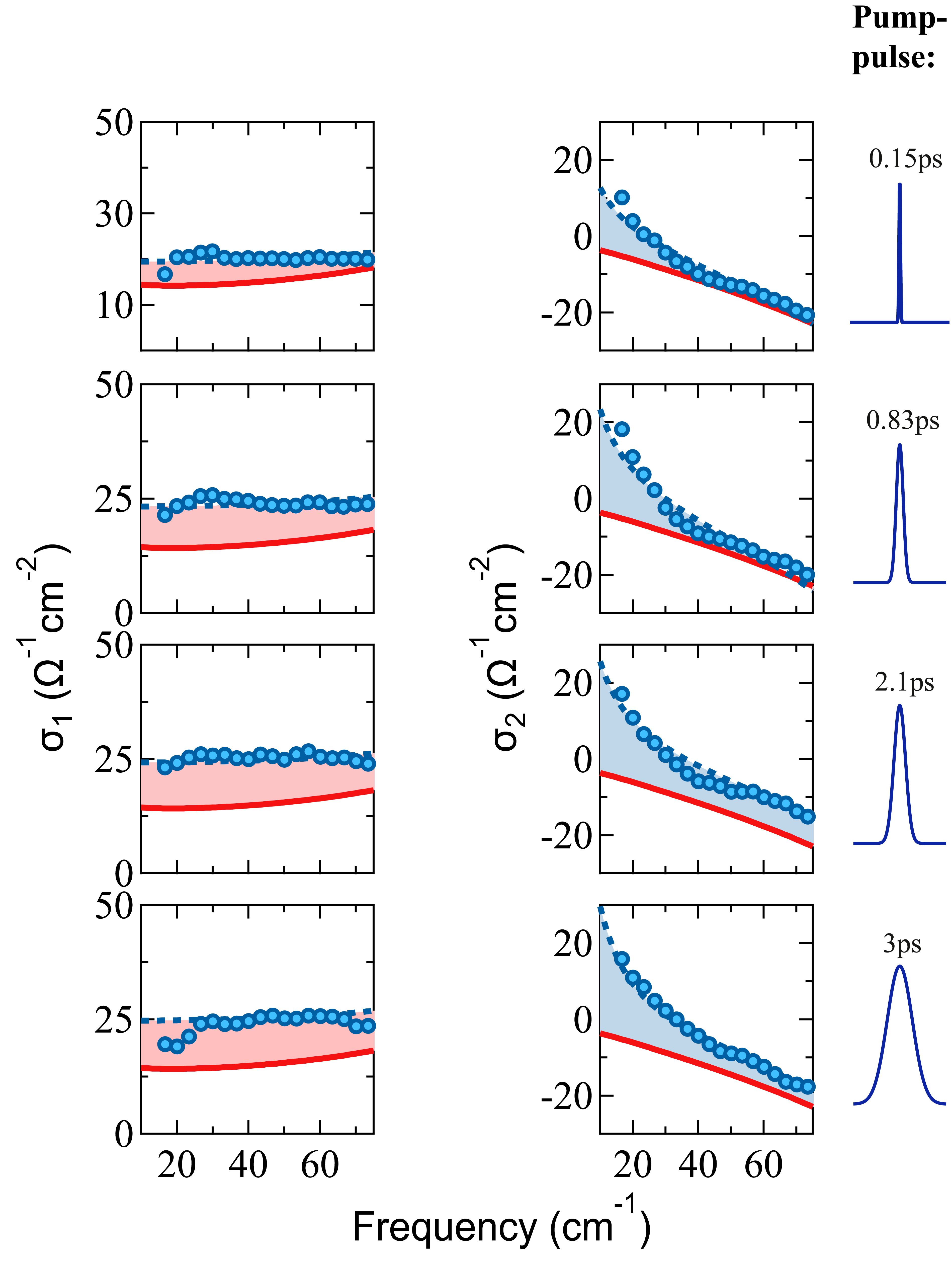}
    \caption{Complex $c$-axis optical conductivity, $\sigma_1(\omega)+i\sigma_2(\omega)$, measured in YBa$_2$Cu$_3$O$_{6.48}$ at $T=100$ K in equilibrium (red) and at one selected pump-probe time delay after photo-excitation (blue circles), corresponding to the peak of the coherent, superconducting-like response. We display data for different mid-infrared pump pulse durations (shown on the right), all taken with the same peak field of 2.2 MV/cm. Dashed blue lines are fits performed with a model including a Josephson plasmon and a Drude term (see Section S4). This dataset was used to produce the color plots in Fig. 2 of the main text.\label{fig:FigS3}}
\end{figure}

\begin{figure}[h]
	\includegraphics[width=1\columnwidth]{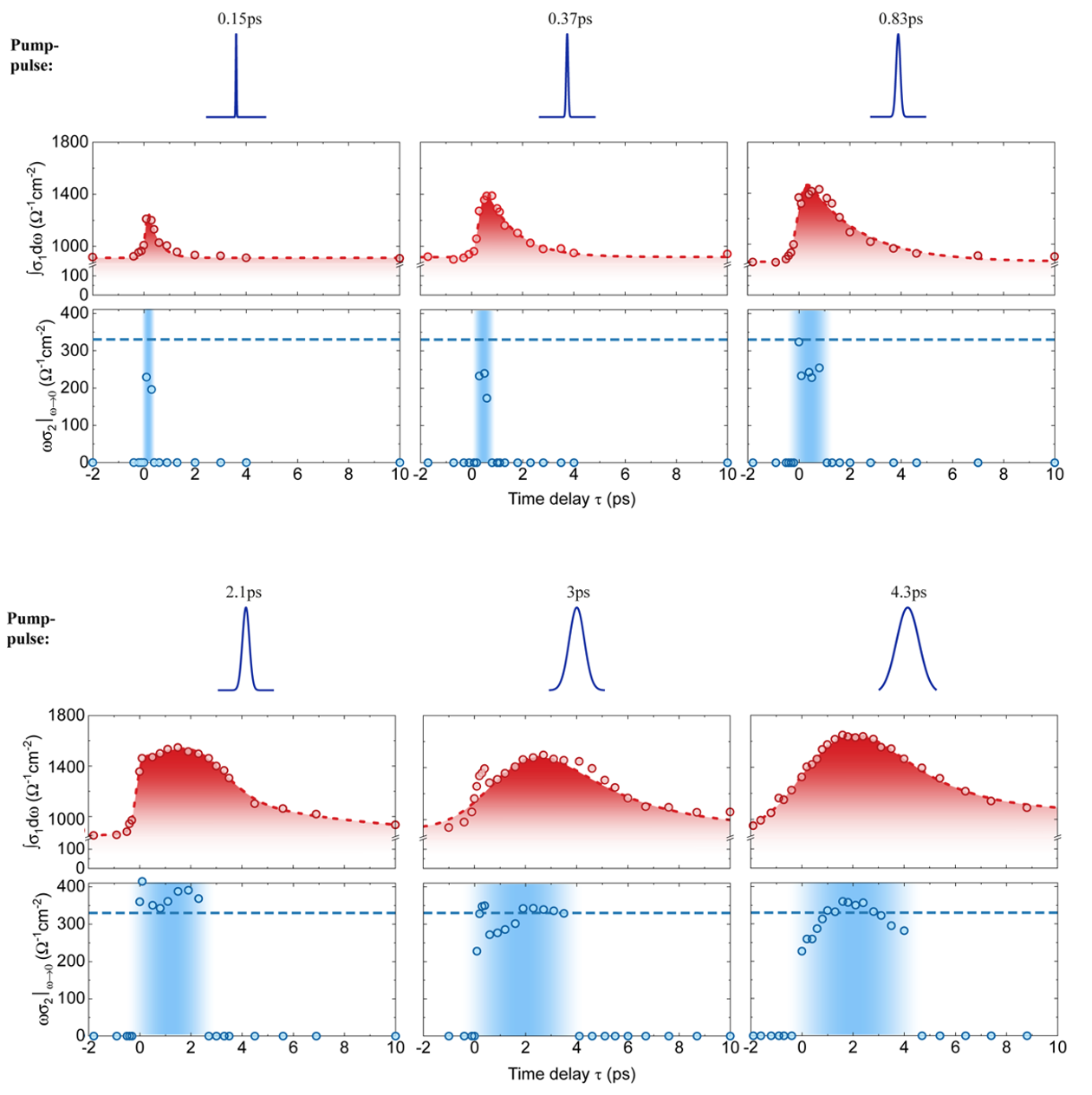}
    \caption{Dynamical evolution of the transient spectral weight, $\int_{20\text{ cm}^{-1}}^{75\text{ cm}^{-1}}\sigma_1(\omega) \mathrm{d}\omega$, indicative of dissipation, and of the coherent superconducting-like response, $\lim_{\omega \to 0} \omega\sigma_2(\omega)$, as a function of pump-probe time delay $\tau$. Data are reported for different pump pulse durations, shown on top of each pair of panels, for a constant peak field of 2.2 MV/cm. The red dashed lines are fits with a finite rise time and an exponential decay, while the blue horizontal lines indicate the 10 K equilibrium value of the superfluid density. The blue shaded areas represent the time delay window for which a finite coherent response was detected. Selected data in this figure were used to produce the plots in Fig. 3 of the main text.\label{fig:FigS4}}
\end{figure}

Figure S3 shows the complex optical conductivity, $\sigma_1(\omega)+i\sigma_2(\omega)$, of YBa$_2$Cu$_3$O$_{6.48}$ at $T=100$ K  in equilibrium (red) and at one selected pump-probe time delay after photo-excitation (blue circles), corresponding to the peak of the coherent response in $\sigma_2(\omega)$ (see also Fig. S4). We report data for different mid-infrared pump pulse durations, all taken with the same peak field of 2.2 MV/cm. The $\sigma_1(\omega)$ spectra show an almost frequency independent increase, indicative of enhanced dissipation, that will reach its maximum for longer time delay (not shown here, see Fig. 2 and Fig. S4). The imaginary conductivity, $\sigma_2(\omega)$, displays instead a $1/\omega$-like divergence at low frequency, which saturates to a given amplitude for pump pulses longer than $\sim 1$ ps.

In Fig. S4 we report the dynamical evolution of $\int_{20\text{ cm}^{-1}}^{75\text{ cm}^{-1}}\sigma_1(\omega) \mathrm{d}\omega$, indicative of dissipation, and $\lim_{\omega \to 0} \omega\sigma_2(\omega)$, which in a superconductor at equilibrium is proportional to the superfluid density, for six different pump pulse durations between 0.15 ps and 4.3 ps. For all data sets, after photo-excitation the $\sigma_1(\omega)$ spectral weight is enhanced and then relaxes on a time scale that clearly exceeds the pump pulse duration. A different evolution is observed instead in $\lim_{\omega \to 0} \omega\sigma_2(\omega)$, which appears to be finite only while the system is driven (blue shading). In this case, excitation with pulses longer than $\sim 1$ ps allows the “zero-temperature” equilibrium superfluid density to be reached (dashed horizonal line).